\documentclass[amsmath,superscriptaddress,twocolumn,showpacs,prb]{revtex4}
\usepackage{bm}
\usepackage[usenames]{color}
\usepackage{graphicx}

\begin{document}

\title{Effect of electron interactions on the conductivity and exchange coupling
energy of disordered metallic magnetic multilayer}
\author{Vladimir A. Zyuzin}
\affiliation{A.F. Ioffe Physical-Technical Institute, 194021
St.Petersburg, Russia} \affiliation{Department of Physics,
University of Utah, Salt Lake City, UT 84112}
\author{A.Yu. Zyuzin}
\affiliation{A.F. Ioffe Physical-Technical Institute, 194021
St.Petersburg, Russia}

\begin{abstract}
We consider the effect of electron-electron interactions on the
current-in-plane (CIP) conductivity and exchange coupling energy of
a disordered metallic magnetic multilayer. We analyze its dependence
on the value of ferromagnetic splitting of conducting electrons and
ferromagnetic layers relative magnetizations orientation. We show
that contribution to the CIP conductivity and exchange coupling
energy as a periodic function of the angle of magnetizations
relative orientation experience $ 2\pi \rightarrow \pi $ transition
depending on the characteristic energies: ferromagnetic splitting of
the conducting electrons and the Thouless energy of paramagnetic
layer.
\end{abstract}

\maketitle

\section{Introduction}

The most interesting features of perfect metallic magnetic
multilayer are the oscillatory behavior of bilinear exchange
coupling energy between ferromagnets \cite{Grun, SSSP} due to
Friedel oscillations, and large magnetoresistance in small magnetic
fields \cite{Baibich, Binasch}.

The magnetic structure of adjoining magnetic layers in perfect
multilayered structure oscillates between ferro and
antiferro-magnetic states with increasing the spacer thickness $L$.
Disorder in layers contributes to biquadratic exchange coupling. It
was shown that fluctuations of thickness of the paramagnetic layer
give rise to biquadratic exchange coupling, often leading to
non-collinear magnetic ordering \cite{Slon}. In
Ref.~\onlinecite{ZySpVagW, Zyuzin} the role of scattering of
conducting electrons by impurities in metallic magnetic multilayer
was studied. It was pointed out that in the case of small (compared
to thickness of the layers) mean free path of conducting electrons,
when average Friedel oscillations are exponentially suppressed, the
exchange coupling energy due to random Friedel oscillations and
correlation effects can have biquadratic form. Transition to the
non-collinear phase in disordered structure with increasing $L$ was
experimentally observed in Ref.~\onlinecite{Fuchs}.

It is established that the magnetoresistance of perfect metallic
magnetic multilayered structure is related to spin-depended
scattering of conducting electrons at the interfaces between the
layers \cite{CamBar, ZhLeF, Nesbet}. In the case of disorder,
scattering of conducting electrons by impurities suppresses the
effect of spin-dependent scattering on magnetoresistance. Moreover,
it was theoretically shown that the magnitude of the
magnetoresistance decreases exponentially in the case of the
current-in-plane (CIP) geometry, when thickness of the nonmagnetic
spacer exceeds the mean free path of conducting electrons
\cite{ZhLeF, BartFert}.

In the present paper we study the effect of electron-electron
interactions on CIP conductivity and exchange coupling energy of
disordered metallic multilayered structure consisting of two
ferromagnetic layers with paramagnetic spacer. We consider the case
when mean free path of conducting electrons is smaller than
thicknesses of layers. It is known that in the disordered conductors
electron-electron interactions result in anomalous contributions to
conductivity, thermodynamic quantities and negative
magnetoconductivity \cite{AlAr, LeeRam}.  Physics behind the effect
of electron-electron interactions in disordered conductors is the
electron scattering by the random Friedel oscillations
\cite{RudinAlGl}. Freidel oscillations in magnetic multilayered
structure do depend on the angle $\varphi$ between directions of
magnetization in magnetic layers. The study of this dependence is
the subject of this paper.

The scattering by Freidel oscillations exists in any type of
ferromagnetic structure: itinerant (\textit{d-} type) or localized
(\textit{f-} type). In what follows, the most relevant factors of
disordered magnetic multilayered structure are the characteristic of
the disorder - the Thouless energy $D/L^{2}$ ($D$ is the conducting
electron diffusion constant), and ferromagnetic splitting of
conducting electrons. Because of that we consider the model of
localized \textit{s-f} magnetism that contains these parameters in
the most transparent way. We propose that each ferromagnetic layer
is described by the homogeneous magnetization, and that directions
of magnetizations of different layers make an angle $\varphi$.

The conductivity and exchange coupling energy in magnetic
multilayered structure are periodic functions of $\varphi$. We show
that depending on the ratio of introduced above characteristic
energies, the angle $\varphi$ dependent contributions of
electron-electron interactions experience the $2\pi \rightarrow \pi$
periodicity transition. Magnitudes of contributions are estimated.

We suppose that our results might be relevant to the series of works
related to transport properties of mesoscopic ferromagnets with
domain walls, where it was experimentally \cite{mesofexp} and
theoretically \cite{yuli, tatara} shown that effects of electron
interactions and weak localization are important.

We would like to mention that results of this paper complement the
theory of electron-electron interactions in disorders conductors
\cite{AlAr, LeeRam} as a study of effect of electron-electron
interactions in spatially inhomogeneous effective magnetic field
that is imposed by the magnetization in layers.

\section{The Model}
The system of our study is disordered metallic magnetic multilayer
structure. It consists of two ferromagnetic layers of the $f$-type
located at $L/2<|z|<L_{1}/2$ and a paramagnetic layer between
($|z|<L/2$). The Hamiltonian of conducting electrons of this system
is:
\begin{equation}
H=H_{0}+H_{C}+H_{sf}.  \label{1}
\end{equation}
Here $H_{0}$ is the Hamiltonian of free electrons in a random field
$V\left( \mathbf{r}\right) $. Treating scattering of free electrons
we use the standard diagram technique \cite{Abrikosov}, which
assumes $\left\langle V\left( \mathbf{r}\right) \right\rangle =0$
and $\left\langle V\left( \mathbf{r}\right) V\left(
\mathbf{r}^{\prime }\right) \right\rangle =1/(2\pi \nu _{0}\tau
)\delta \left( \mathbf{r}-\mathbf{r}^{\prime }\right) $. Here $\nu
_{0}$ is the density of states at the Fermi level per one spin,
$\tau $ is the electron mean free time, and we have set $\hbar=1$.
We assume the Hamiltonian of free electrons to be the same in every
layer.

The second term $H_{C}$ is the Hamiltonian of the Coulomb
interaction between conducting electrons.
$H_{C}=\frac{1}{2}\int\Psi^{\dag}_{\alpha}({\bf
r})\Psi^{\dag}_{\beta}({\bf r}^{\prime})\frac{e^2}{|{\bf r}-{\bf
r}^{\prime}|}\Psi_{\beta}({\bf r}^{\prime})\Psi_{\alpha}({\bf
r})d{\bf r}d{\bf r}^{\prime}$, where $ \Psi _{\alpha }^{\dag}\left(
\mathbf{r}\right) $ and $\Psi _{\beta }\left( \mathbf{r}\right) $
are the electron creation and annihilation operators
correspondingly. We will treat Coulomb interaction within the random
phase approximation.

The third term $H_{sf}$ is the Hamiltonian of $s-f$ exchange in the
ferromagnetic layers. At temperatures much lower than the Curie
temperature one might neglect electron-magnon interaction, therefore
\begin{eqnarray}
H_{sf} = &I\sum\limits_{i} \Psi _{\alpha }^{\dag}\left(
\mathbf{r_{i}}\right) (\mathbf{S_{i}}\cdot{\bm \sigma ~}_{\alpha
\beta
})\Psi _{\beta }\left( \mathbf{r_{i}}\right)\rightarrow  \nonumber\\
& ISn_{S}\int\limits_{F}{d\mathbf{r~}\Psi _{\alpha }^{\dag}}\left(
\mathbf{r}\right) ({\mathbf{n~}\left( \mathbf{r}\right)\cdot{\bm
\sigma ~}_{\alpha \beta })\Psi _{\beta }}\left( \mathbf{r}\right)
\end{eqnarray}
Here  $S_{i}$ is a spin of localized $f$-electrons, $n_{S}$ is their
density. $I$ is the $s-f$ exchange interaction. $\mathbf{n}\left(
\mathbf{r}\right)$ is the ferromagnetic layer magnetization
direction unit vector. Integration here is over the ferromagnetic
layers.

Neglecting the contribution of electron-magnon interaction to the
conductivity and exchange coupling energy, we assume that the
Coulomb energy per electron is larger than the ferromagnetic
splitting $ISn_{S}$.

Finally we consider the following Hamiltonian of a disordered
metallic magnetic multilayer:
\begin{equation}
H=H_{0}+H_{C}+\frac{\epsilon _{exc}}{2}\int\limits_{F}{d {\bf
r~}\Psi _{\alpha }^{\dag}}\left( \mathbf{r}\right) ({{\bf
n~}(z)\cdot{\bm \sigma ~}_{\alpha \beta })\Psi _{\beta }}\left(
\mathbf{r}\right). \label{finham}
\end{equation}
This Hamiltonian describes a Fermi liquid in an effective
inhomogeneous magnetic field that acts only on electron spin. This
magnetic field results in the spin splitting with energy
$\epsilon_{exc}=2ISn_{S}$ in the ferromagnetic layers, and in zero
splitting in the paramagnetic layer. Direction of this magnetic
field is defined by the ferromagnetic layer magnetization direction
unit vector $\mathbf{n}(z)$: $\mathbf{n}\left( z\right)
=\mathbf{n}_{1}\ $at $ -L_{1}/2<z<-L/2 $ and $\mathbf{n}\left(
z\right) =\mathbf{n}_{2}\ $at $L_{1}/2>z>L/2$, with
$(\mathbf{n}_{1}\cdot\mathbf{n}_{2})=cos\varphi$.

Now we are ready to study the effect of electron interaction on the
CIP conductivity and exchange coupling energy between the
ferromagnetic layers. We will be interested in its dependence on the
angle between the direction of magnetization $\varphi$,
ferromagnetic splitting energy $\epsilon_{exc}$, and paramagnetic
thickness $L$.

\subsection{Conductivity}
We use Kubo linear response formalism in order to study the effect
of electron interaction on conductivity. Treating electron
interaction as a perturbation, we come up with corrections to
conventional Drude formula for conductivity. We study only the first
order perturbation correction. The magnetic field dependent, or in
our case magnetization dependent, correction to conductivity comes
from the interaction between two electrons in the triplet state
(so-called the Hartree-type correction) \cite{AlAr, LeeRam}.
Disorder averaged correction can be described with the help of a
Feynman diagram, shown in the fig. $1A$. Black squares in the figure
stand for the diffusion ladder that represents electron`s charge and
spin densities propogation.
\begin{figure}[]
\centering  \includegraphics[width=8.5cm]{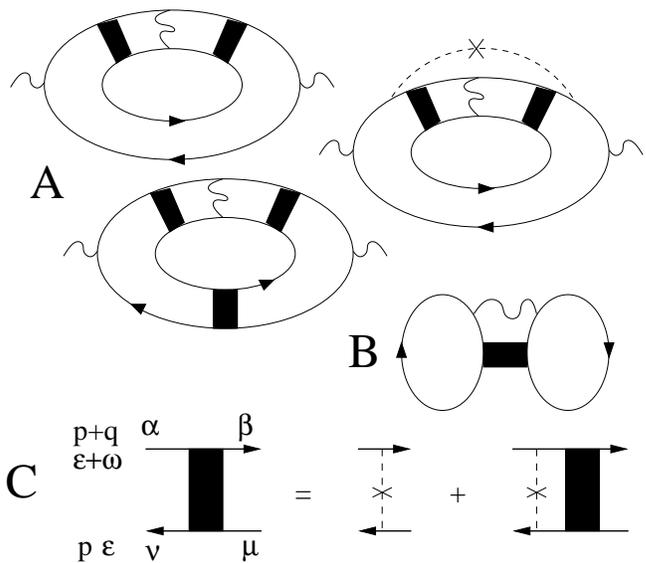} \caption{A.
Feynman diagram representation of the Hartree-type corrections to
conductivity. The solid lines in the diagrams denote the disorder
averaged Green`s functions, wavy line presents the screened Coulomb
interaction, dotted line with the cross represents impurity
scattering, and the black square stands for the diffusion ladder. B.
Hartree-type correction to thermodynamic potential. C. Equation for
the diffusion ladder}
\end{figure}
Calculations of presented diagrams for the homogeneous magnetic
field are carried out in Ref.~\onlinecite{AlAr}. We make use of the
bilinear representation of the diffusion ladder in order to
calculate these diagrams for inhomogeneous magnetic field. Our
calculations show that in the case of the CIP geometry all diffusion
ladder combinations that appear in the figure can be simplified in
to the following expression:
\begin{eqnarray}  \label{cip}
&\delta\sigma\left( z\right) = i F(z)\frac{e^{2} }{8\pi^{2}}
\nonumber \\
&\times\int\limits_{-\infty}^{\infty}d\omega\frac{d}{d\omega}(\omega\coth (\frac{%
\omega}{2T}))Tr{\cal D}_{\mu\mu}^{\alpha\alpha}(z,z,0,-i\omega),
\label{2}
\end{eqnarray}
where $F\left( z\right) =F_{fer}\theta\left( \left\vert z\right\vert
-L/2\right)+F_{par}\theta\left( L/2-\left\vert z\right\vert \right)
$ is the electron interaction constant that is discussed in
Ref.~\onlinecite{AlAr}. We assume that interaction constants might
be different in ferromagnetic and paramagnetic layers. $T$ is the
temperature, ${\cal D}_{\mu\nu}^{\alpha\beta}$ is the diffusion
ladder. For the CIP geometry it is convenient to consider
conductance $\delta G\equiv\int dz\delta\sigma\left( z\right) $ as a
sum over the paramagnetic and ferromagnetic layers.

\subsection{Exchange coupling energy}

Contribution to the multilayer thermodynamic potential, which
depends on the relative orientation of the ferromagnetic
magnetizations, is also due to interaction between two electrons in
the triplet state. This correction is presented in the fig. $1B$ and
given by an expression \cite{AlAr}
\begin{eqnarray}\label{coupling}
&\frac{\delta \Omega \left( \varphi \right) }{S}=&  \\
&\frac{T}{4}\sum\limits_{\left\vert \omega _{n}\right\vert \tau
<1}\left\vert \omega _{n}\right\vert \int
\frac{d^{2}\mathbf{q}}{\left( 2\pi \right) ^{2}}\int dzF\left(
z\right) {Tr}{\cal D}_{\mu \mu }^{\alpha \alpha }\left(
z,z,\mathbf{q},\omega _{n}\right),   \nonumber
\end{eqnarray}
where $\omega _{n}=2\pi nT$ is the Matsubara frequency, $S$ is the
area of the system. The angle-dependent part of $\delta \Omega
\left( \varphi \right) $ determines exchange coupling energy of
adjoint ferromagnetic layers.

\subsection{Diffusion ladder}

The angle $\varphi$ dependence in expressions (\ref{cip}) and
(\ref{coupling}) is due to the diffusion ladder. Graphical equation
for the diffusion ladder ${\cal D}_{\mu \nu }^{\alpha \beta }\left(
z,z^\prime,\mathbf{q},\omega _{n}\right)$ is shown in the fig. $1C$,
and in the case of diffusion approximation ($|\omega_n|\tau <1$,
$\epsilon_{exc}\tau <1$, and $|q|v_{F}\tau<1$ with $v_{F}$ being the
Fermi velocity) satisfies the following differential equation
\begin{eqnarray}\label{dif}
&\left( -D\frac{d^{2}}{dz^{2}}+Dq^{2}+|\omega _{n}|\right) {\cal
D}_{\mu \nu
}^{\alpha \beta }  \nonumber  \label{eq4} \\
&+i\frac{\epsilon _{exc}}{2}\mathbf{n~}(z)\left( {\bm \sigma}
_{\alpha \gamma }{\cal D}_{\mu \nu }^{\gamma \beta }-{\cal D}_{\mu
\gamma }^{\alpha \beta }{\bm \sigma} _{\gamma \nu }\right)
sign\omega _{n}=\\
&\delta (z-z^{\prime })\delta _{\alpha \beta }\delta _{\mu \nu
},\nonumber
\end{eqnarray}
Here $D=\frac{1}{3}v_{F}^{2}\tau$ is the diffusion constant. We
assume it to be the same in every layer. We notice that in
expressions (\ref{cip}) and (\ref{coupling}) only trace of the
diffusion ladder appears. Since the trace is invariant under unitary
transformations, we may choose the ferromagnetic layers
magnetization direction unit vector in the convenient for further
calculations: in $-L_{1}/2<z<-L/2$ the direction is
$\mathbf{n}_{1}=(1,0,0)$, and in $L_{1}/2>z>L/2$ it is
$\mathbf{n}_{2}=(\cos \varphi ,0,\sin \varphi )$.

Solution of the equation (\ref{dif}) has to satisfy the boundary
conditions $\frac{d}{dz}{\cal D}_{\mu \nu }^{\alpha \beta
}(z,z^{\prime }, \mathbf{q},\omega _{n})=0$ at $z=\pm L_{1}/2$, and
the continuity conditions at $z=\pm L/2$.

\section{$2\protect\pi\rightarrow\protect\pi$ transition}

Here we present our main results by skipping all derivations that
will be given in the appendix to the paper. We consider only
angle-dependent parts of (\ref{cip}) and (\ref{coupling}).

Characteristic energies of a magnetic multilayer are the Thouless
energy of the paramagnetic layer $D/L^{2}$ and ferromagnetic
splitting energy $\epsilon _{exc}$. We present results for the cases
of large ferromagnetic thickness $d=L_{1}/2-L/2$: $\epsilon
_{exc}d^{2}/D>1$ and for small temperatures: $\epsilon _{exc}>>T$.

Our main finding is the following: depending on the ratio of the
Thouless energy and ferromagnetic splitting energy, contribution to
the CIP conductance (\ref{cip}) and exchange coupling energy between
ferromagnetic layers (\ref{coupling}) might have $2\pi $ or $ \pi $
periodicity as a function of $\varphi$. $2\pi \rightarrow \pi $
transition occurs when ratio $\epsilon _{exc}L^{2}/D$ increases.

(i) In the case of large Thouless energy $\epsilon _{exc}L^{2}/D<1$
the leading term the in angle-dependent parts of the CIP conductance
and exchange coupling energy is proportional to $\cos {\varphi }$.
One can show that ferromagnetic layers give the main contribution to
the conductance as well as to exchange coupling energy:
\begin{equation}
\delta G\simeq -F_{fer}\frac{e^{2}}{32\pi ^{2}}\cos \varphi ,
\label{5a}
\end{equation}
\begin{equation}
\frac{\delta \Omega \left( \varphi \right) }{S}\simeq
-\frac{\epsilon _{exc}^{2}\cos \varphi }{2\left( 4\pi \right)
^{2}D}\left( 0.06F_{par}+0.7F_{fer}\ln \left( \frac{D}{\epsilon
_{exc}L^{2}}\right) \right).   \label{6a}
\end{equation}
Correction to conductance originating from the paramagnetic layer is
reduced by a factor of $\sqrt{\epsilon _{exc}L^{2}/D}$.

(ii) In the limit of $\epsilon _{exc}L^{2}/D>1$ the leading term in
(\ref{cip}) and (\ref{coupling}) is $\cos ^{2}{\varphi }$, and it is
independent on the ferromagnetic splitting energy $\epsilon _{exc}$:
\begin{equation}
\delta G\simeq -\left( 0.7F_{fer}+0.03F_{par}\right) \frac{e^{2}}{4\pi ^{2}}%
\cos ^{2}\varphi ,   \label{7}
\end{equation}%
\begin{equation}
\frac{\delta \Omega \left( \varphi \right) }{S}\simeq \frac{\left(
3F_{fer}+F_{par}\right) }{2\left( 4\pi \right) ^{2}D}\left( \frac{D}{L^{2}}%
\right) ^{2}\cos ^{2}\varphi .   \label{8}
\end{equation}

The physics behind this transition is the following. According to
Ref.~\onlinecite{AlAr} the magnetic field dependent contributions to
the conductivity and exchange coupling energy are due to interaction
of electrons with total spin $J=1$ and projection $J_{z}=\pm 1$. In
our case when direction of $\mathbf{n}\left( \mathbf{r}\right)$
varies in space, projections $J_{z}=0,\pm 1$ are defined only
locally.

At small $\epsilon _{exc}$ the leading contribution is due to
interaction of electrons with total spin $J=1$ and projections
$J_{z}=\pm 1$, and the dependence of the conductivity and of the
exchange coupling energy is proportional to $\cos{\varphi }$. In the
case of large $\epsilon _{exc}$ diffusion modes with total spin
$J=1$ and projection $J_{z}=\pm 1$ do not penetrate into
ferromagnetic layers and their contributions to the conductivity and
the exchange coupling energy are suppressed. The main contribution
in this case is due to the diffusion mode with $J=1$, $J_{z}=0$ and
proportional to $\cos ^{2}{\varphi }$. Let us note, that in this
case the electron-magnon interaction, which creates $J_{z}=\pm 1$
diffusion modes in ferromagnetic layers might be neglected even if
the Coulomb energy per electron is smaller than the ferromagnetic
splitting.

At temperatures $T>D/L^{2}$ the contribution to the conductance
decreases as
\begin{equation}
\delta G=-\frac{e^{2}}{4\pi ^{2}}\frac{0.2L^{2}}{DT}\left[ F_{par}+5F_{fer}
\right] \cos ^{2}\varphi .  \label{9a}
\end{equation}
And exchange coupling energy decreases with the temperature as
$exp(-L\sqrt{8\pi T/D})$.

In the Fermi liquid theory the interaction constants are considered
as some parameters. In the case of weakly non ideal 3D Fermi gas
$F=\kappa^2/ (2p_{F}^2) ln(1+4p_{F}^2/\kappa^2)$ , where $\kappa$
and $p_{F}$ are the inverse screening length and the Fermi momentum,
correspondingly \cite{AlAr}. For the Coulomb interaction $F_{par}$
and $F_{fer}$ are both positive. In both limits, (i) and (ii),
contribution to conductance has a minimum at $\varphi =0$. Exchange
coupling energy has a minimum at $ \varphi=0$ in the first limit,
and at $\varphi=\pi/2$ in the second limit.

Let us estimate the value of exchange coupling energy. Taking $D=10
cm^2/sec$, $\epsilon _{exc}=300K\times k_{B}$ (where $k_{B}$ is the
Boltzmann constant), we obtain that crossover occurs at the
paramagnetic layer thickness of $L=\sqrt{\hbar D/\epsilon
_{exc}}=5nm$. Plugging these values into equations (\ref{6a}) and
(\ref{8}), and assuming $F_{fer}\sim F_{par}\sim 1$ we obtain
$\delta \Omega \left( \varphi \right) /S \sim 0.5 \times 10^{-3}
erg/cm^2$.

Correction to conductance is of the order of percents of
$\frac{e^2}{\hbar}$.

\section{Conclusion}

In the present paper we have considered properties of a disordered
magnetic metallic multilayer in the case of small electron mean free
path compared to paramagnetic layer thickness. The angle $\varphi$
dependence of the CIP conductance and of the exchange coupling
energy is determined by the electron-electron interactions.
Depending on the ratio of the Thouless and ferromagnetic splitting
energies these quantities might be $2\pi$ as well as $\pi$ periodic
functions of $\varphi$.

Obtaining our results we have neglected the effect of spin-orbit
scattering. This scattering smears out the magnetic dependence of
electron interaction corrections to thermodynamic and kinetic
properties\cite{AlAr}. However, this effect become crucial for
paramagnetic thickness (larger than the spin relaxation length due
to spin-orbital scattering) when values of calculated quantities
already become too small.

We would like to point out that in Ref.~\onlinecite{Zyuzin}
contribution to exchange coupling energy due to the interaction was
considered only in the paramagnetic layer. In this paper we have
shown that the contribution of ferromagnetic layers is larger than
that of paramagnetic layer.

\bigskip
\begin{acknowledgements}
This work is supported by the Russian Fund for Fundamental Research
grants 05-02-17816 and 06-02-17047.
\end{acknowledgements}

\appendix
\section{Derivations}

Here we describe the main steps of derivation. We solve the
differential equation (\ref{dif}) for diffusion ladder. According to
the expressions (\ref{cip}), (\ref{coupling}) the solutions of
interest are those with $z$ and $z^{\prime }$ lying in the same
layer.

(i) Let us first derive the diffusion ladder when $z^{\prime }<-L/2$
is in ferromagnetic layer. It is convenient to present solution in
paramagnetic layer ($\left\vert z\right\vert <L/2$) in the following
form:
\begin{equation}\label{diffpara1}
{\cal D}_{\mu \nu }^{\alpha \beta }=A_{\mu \nu }^{\alpha \beta
}e^{-Qz}+U_{\alpha \gamma }^{\dag}B_{\mu \delta }^{\gamma \beta
}U_{\delta \nu }e^{Qz}.
\end{equation}
And in ferromagnetic layer at $-L_{1}/2<z^{\prime }<z<-L/2$ solution
that satisfies the boundary condition at $z=-L_{1}/2$ is convenient
to present in the following form:
\begin{eqnarray}\label{diffferro1}
&{\cal F}_{\mu \nu }^{\alpha \beta }={\cal F}_{0\mu \nu }^{\alpha
\beta }+\sum_{\pm }P_{\pm }^{\alpha \gamma }C_{\mu \lambda }^{\gamma
\beta }P_{\pm }^{\lambda
\nu }\cosh Q\left( z+\frac{L_{1}}{2}\right)   \nonumber \\
&+P_{+}^{\alpha \gamma }E_{\mu \lambda }^{\gamma \beta
}P_{-}^{\lambda \nu
}\cosh \left( Q_{1}\left( z+\frac{L_{1}}{2}\right) \right)   \\
&+P_{-}^{\alpha \gamma }M_{\mu \lambda }^{\gamma \beta
}P_{+}^{\lambda \nu }\cosh \left( Q_{1}^{\ast }\left(
z+\frac{L_{1}}{2}\right) \right) ,  \nonumber
\end{eqnarray}
where
\begin{eqnarray}
&{\cal F}_{0\mu \nu }^{\alpha \beta }= \\
&\frac{\sum_{\pm }P_{\pm }^{\alpha \beta }P_{\pm }^{\mu \nu
}}{DQ}\exp \left( -Q(\frac{L_{1}}{2}+z)\right) \cosh Q\left(
z^{\prime }+\frac{L_{1}}{2}\right)
\nonumber \\
&+\frac{P_{+}^{\alpha \beta }P_{-}^{\mu \nu }}{DQ_{1}}\exp \left(
-Q_{1}(\frac{L_{1}}{2}+z)\right) \cosh Q_{1}\left( z^{\prime
}+\frac{L_{1}}{2}\right) +
\nonumber \\
&+\frac{P_{-}^{\alpha \beta }P_{+}^{\mu \nu }}{DQ_{1}^{\ast }}\exp
\left( -Q_{1}^{\ast }(\frac{L_{1}}{2}+z)\right) \cosh Q_{1}^{\ast
}\left( z^{\prime }+\frac{L_{1}}{2}\right) ,  \nonumber
\end{eqnarray}%
where $P_{\pm }=(1\pm \sigma _{z})/2$ are projectors on the spin up
and down states correspondingly;
 $Q=\sqrt{\mathbf{q}^{2}+\left\vert \omega _{n}\right\vert /
D}$, $Q_{1}=\sqrt{\mathbf{q}^{2}+(\left\vert \omega _{n}\right\vert
+i\epsilon _{exc}sign\omega _{n})/D}$, and $U=\exp(i\varphi \sigma _{y}/2)$ is the matrix of rotation along the $%
y$ axis. $A,B,C,E,M$ are matrixes-constants to be determined with
the help of continuity conditions.

It is convenient to obtain closed set equations for matrixes $A\pm
B$ with the help of the continuity conditions for rotated diffusion
ladder (\ref{diffpara1}) $\widehat{{\cal D}}_{\mu\nu}^{\alpha\beta}\equiv U_{\alpha\gamma}%
{\cal D}_{\mu\lambda}^{\gamma\beta}U^{\dag}_{\lambda\nu}$ at
$z=L/2$:
\begin{eqnarray}\label{bc1}
&{\hat {\cal R}}(Q)P_{\pm}^{\alpha\gamma} \widehat{{\cal
D}}_{\mu\lambda}^{\gamma\beta}P_{\pm}^{\lambda\nu}={\hat {\cal
R}}(Q_1)P_{+}^{\alpha\gamma} \widehat{{\cal
D}}_{\mu\lambda}^{\gamma\beta}P_{-}^{\lambda\nu}=\nonumber\\
&{\hat{\cal R}}(Q^{\ast}_1)P_{-}^{\alpha\gamma} \widehat{{\cal
D}}_{\mu\lambda}^{\gamma\beta}P_{+}^{\lambda\nu}=0,
\end{eqnarray}
where we have used the differential operator ${\hat {\cal
R}}(Q)=\left( Q\sinh\left( \frac{Q(L-L_{1})}{2}\right) -\cosh\left(
\frac{Q(L-L_{1})}{2}\right) \frac{d}{dz}\right).$ And continuity
conditions for diffusion ladder (\ref{diffpara1}) at $z=-L/2$ are
\begin{eqnarray}
{\hat {\cal L}}(Q) P_{\pm}^{\alpha\gamma}%
{\cal D}_{\mu\lambda}^{\gamma\beta}P_{\pm}^{\lambda\nu}  &
=\frac{P_{\pm}^{\alpha\beta}P_{\pm
}^{\mu\nu}}{D}\cosh Q\left(  z^{\prime}+\frac{L_{1}}{2}\right) ,\nonumber\\
{\hat {\cal L}}(Q_1) P_{+}^{\alpha\gamma}{\cal
D}_{\mu\lambda}^{\gamma\beta}P_{-}^{\lambda\nu} &
=\frac{P_{+}^{\alpha\beta}P_{-}^{\mu\nu}}{D}\cosh Q_{1}\left(  z^{\prime}+\frac{L_{1}}{2}\right) ,\nonumber\\
{\hat {\cal L}}(Q^{\ast}_1) P_{-}^{\alpha\gamma} {\cal
D}_{\mu\lambda}^{\gamma\beta}P_{+}^{\lambda\nu}  &  =\frac
{P_{-}^{\alpha\beta}P_{+}^{\mu\nu}}{D}\cosh Q_{1}^{\ast}\left(
z^{\prime }+\frac{L_{1}}{2}\right),\nonumber
\end{eqnarray}
with ${\hat {\cal L}}(Q)=\left( Q\sinh\left(
\frac{Q(-L+L_{1})}{2}\right) -\cosh\left(
\frac{Q(-L+L_{1})}{2}\right)  \frac{d}{dz}\right)$.

Solving for matrixes $A,B$ and subtracting angle independent part we
obtain the trace of the diffusion ladder for ferromagnetic layer
(\ref{diffferro1})
\begin{widetext}
\begin{eqnarray}\label{ferr}
&\int\limits_{|z|>L/2}Tr{\cal
F}_{\mu\mu}^{\alpha\alpha}(z,z,\mathbf{q},\left\vert
\omega_{n}\right\vert)dz
=\nonumber\\
&  \frac{-4\Lambda_{1}}{Q}\left[  \frac{\operatorname{Re}\left(
1+\Lambda\right)  \left[  \Lambda^{\ast}+\cos\varphi\right]
}{W_{+}}+\left(
\Lambda\rightarrow-\Lambda,\Lambda_{1}\rightarrow-\Lambda_{1}\right)
\right] \frac{1}{D}\left[  \frac{\sinh\left(  2Qd\right)
}{Q}+2d\right]  \nonumber\\& +\frac{8Q}{D}\operatorname{Re}\left[
\frac{2+\Lambda+\Lambda\Lambda _{1}+\left(
2\Lambda_{1}+\Lambda\Lambda_{1}+\Lambda\right)  \cos\varphi
}{2Q_{1}\left(  Q^{2}-Q_{1}^{2}\right)  W_{+}}+\left(
\Lambda\rightarrow
-\Lambda,\Lambda_{1}\rightarrow-\Lambda_{1}\right)  \right],
\end{eqnarray}
\end{widetext}
where we have used next notations:
\begin{eqnarray}
&\Lambda=\frac{Q_{1}\sinh (dQ_{1})-Q\cosh (dQ_{1})}{Q_{1}\sinh
(dQ_{1})+Q\cosh (dQ_{1})}\exp\left(  -QL\right),\nonumber\\
&W_{\pm}=\operatorname{Re}\left(  1\pm\Lambda^{\ast}\right)
\left[1+\Lambda\Lambda_{1}\pm\left(  \Lambda_{1}+\Lambda\right)
\cos\varphi\right],\nonumber\\
&\Lambda_{1}=\exp\left(-QL_{1}\right).\nonumber
\end{eqnarray}

(ii) The same procedure can be applied for the case when
$z^{\prime}$ is in paramagnetic layer. Now the solution of equation
$(\ref{eq4})$ when $z$ is in paramagnetic layer is convenient to
present in the following form:
\begin{equation}\label{diffpara2}
{\cal
D}_{\mu\nu}^{\alpha\beta}=A_{\mu\nu}^{\alpha\beta}e^{-Qz}+U_{\alpha\gamma}
^{\dag}B_{\mu\delta}^{\gamma\beta}U_{\delta\nu}e^{Qz}+\frac{e^{-Q|z-z^{\prime}|}
}{2QD}\delta_{\alpha\beta}\delta_{\mu\nu}.
\end{equation}
In this case the continuity conditions for the rotated diffusion
ladder (\ref{diffpara2}) $\widehat{{\cal D}}_{\mu\nu}^{\alpha\beta}\equiv U_{\alpha\gamma}%
{\cal D}_{\mu\lambda}^{\gamma\beta}U^{\dag}_{\lambda\nu}$ at $z=L/2$
are the same as (\ref{bc1}). At $z=-L/2$ continuity conditions for
the diffusion ladder (\ref{diffpara2}) change as
\begin{eqnarray}
&{\hat {\cal L}}(Q)P_{\pm}^{\alpha\gamma}%
{\cal D}_{\mu\lambda}^{\gamma\beta}P_{\pm}^{\lambda\nu}={\hat {\cal
L}}(Q_1) P_{+}^{\alpha\gamma}{\cal
D}_{\mu\lambda}^{\gamma\beta}P_{-}^{\lambda\nu}
=\nonumber\\
&{\hat {\cal L}}(Q^{\ast}_1) P_{-}^{\alpha\gamma} {\cal
D}_{\mu\lambda}^{\gamma\beta}P_{+}^{\lambda\nu}  =0 .\nonumber
\end{eqnarray}
Solving for matrixes $A,B$ and subtracting angle independent part we
obtain the diffusion ladder for paramagnetic layer (\ref{diffpara2})
, whose trace is
\begin{widetext}
\begin{eqnarray}\label{para}
&\int\limits_{|z|<L/2}Tr{\cal D}_{\mu\mu}^{\alpha\alpha}(z,z,\mathbf{q},\left\vert \omega_{n}\right\vert)dz= \nonumber\\
&  -\frac{L\operatorname{Re}\Lambda^{\ast}\left(  1+\Lambda\right)
}{2DQW_{+}}\left[  \left(  \cos\varphi+\Lambda_{1}\right)  +\left(
1+\Lambda_{1}\cos\varphi\right)  \frac{\sinh{QL}}{QL}\right] -\frac{L\operatorname{Re}\Lambda^{\ast}\left(  1-\Lambda\right)  }%
{2DQW_{-}}\left[  \left(  -\cos\varphi+\Lambda_{1}\right)  +\left(
1-\Lambda_{1}\cos\varphi\right)  \frac{\sinh{QL}}{QL}\right]   \nonumber\\
&  +\Lambda_{1}L\frac{\operatorname{Re}\left[  \left(
1+\Lambda^{\ast }\right)  \left(  1+\Lambda\cos\varphi\right)
\right]  \frac{\sinh{QL}}{QL}-\operatorname{Re}\left[  \left(
1+\Lambda^{\ast}\right) \left( \Lambda+\cos\varphi\right)  \right]
}{2DQW_{+}} +\Lambda_{1}L\frac{\operatorname{Re}\left[  \left(
1-\Lambda^{\ast }\right)  \left(  1-\Lambda\cos\varphi\right)
\right] \frac{\sinh{QL}}{QL}-\operatorname{Re}\left[  \left(
1-\Lambda\right)  \left(  \Lambda^{\ast
}-\cos\varphi\right)  \right]  }{2DQW_{-}}.%
\end{eqnarray}
\end{widetext}

The results (\ref{5a})$-$(\ref{9a}) will be recovered if one will
plug diffusion ladder traces (\ref{ferr}) and (\ref{para}) into the
expressions for the conductivity (\ref{cip}) and exchange coupling
energy (\ref{coupling}). The integrals over the frequency in
resulting expressions can be evaluated as following: in the case of
$\epsilon_{exc}L^{2}/D,\epsilon_{exc}d^{2}/D>1$ the main
contribution is due to frequencies $\left\vert \omega_{n}\right\vert
<\epsilon_{exc}$ where
$\Lambda\simeq\operatorname{Re}\Lambda\simeq\exp\left( -QL\right)$.
In this case expressions (\ref{cip}) and (\ref{coupling}) are even
functions of $cos{\varphi}$. In region from $\epsilon_{exc}L^{2}/D$
$<1$ the main contribution is due to $\left\vert
\omega_{n}\right\vert
> \epsilon_{exc}$ in this case $Re\Lambda\sim|\Lambda|^{2}$, and
expressions (\ref{cip}) and (\ref{coupling}) have the 2$\pi$
periodicity.

\end{document}